# Particles, Localization and "Collapse"

Anthony Rizzi
Institute for Advanced Physics, arizzi@iapweb.org [1]

**Abstract:** Hegerfeldt's theorem shows that the localization of a particle (or field) is not possible under very general assumptions of quantum mechanics and special relativity. This theorem, and further developments and associated theorems force penetrating questions on the proper understanding of the theory and its use. In particular, the use of the collapse postulate is called into question. The existence of particles and even localized fields are called into question. And, this in turn challenges local measurements. It generally forces one to consider the profound problems and even wrong results that arise from the collapse postulate. We give a qualitatively new example of such a wrong result. Researchers and students alike are bothered by these fundamental issues. Students ability to learn the material, and even their interest in physics, can be compromised by such disturbing conundra that seem to make physics, or at least QFT and its QM limit, seem incomprehensible and even inconsistent. This article resolves these issues in a way accessible to undergraduates, by applying the natural (minimal) interpretation of quantum mechanics (QM) (the ensemble interpretation) and explicating and clarifying the basic principles and premises used in QM and the Hegerfeldt theorem and its application to measurements.

## Introduction

The Copenhagen or "the collapse postulate"[2] interpretation is the most popularly assumed interpretation of quantum mechanics, including in the way it is taught in textbooks.[3] This paper brings out how the collapse postulate is inconsistent with limits on localization inherent in ordinary quantum mechanics (QM or OQM). We draw out the unreasonableness of the wavefunction collapse postulate, which, in turn, implies a physical collapse.[4] Through the vehicle of the localization problem revealed by the Hegerfeldt theorem as it appears in the two slit problem, it will be shown that, despite being largely unknown (or even addressed) in the community and the literature, collapse is, in fact, merely a sometimes convenient approximation used in a natural interpretation of QM, not a profound postulate. This understanding simplifies the presentation of quantum mechanics, while it frees it from many intractable problems.[5] If taken, we no longer have to consider the Schrödinger cat that is in a superposition of dead and alive when we are not measuring! We can answer "NO!" to Einstein's quip: "do you really think the moon is not there when you're not looking?"[6] We no longer to need to hold onto the contradiction of existing and not existing. No need to postulate "many worlds" (MW) to avoid it, nor need we avert to MW to preserve the fact that the only kind of quantum state evolution is the Schrodinger equation (SE). No need to postulate all kinds of extras.[7] The ensemble (minimal) interpretation comes naturally to the fore in confronting the problem of collapse and localization.

[1] This paper was written in 2024, vetted in 2025 and early 2026.
[2] Also called the projection postulate.
[3] This is despite the fact that theorists seem to more heavily lean towards supporting the many worlds when asked for their preferred interpretations.
[4] Note, for those who have not focused much on the issue, let me emphasize that collapse is not the idea that, after measurements, one can make an approximation that the wavefunction is "localized" to one state, i.e. that one can ignore parts of the wavefunction (state) that aren't relevant to one's particular problem at one's particular scales. Such approximations are valid in certain circumstances and if they are properly done. In contrast, "collapse," as we will see, is the idea that the wavefunction *collapses* to an eigenstate upon measurement, thereby violating the Schrodinger Equation evolution. In all this, one must be careful not to let "it works" (see fn 72) substitute for understanding why it works. Also, why it works helps us know when it works!

Hegerfeldt's theorem[8] shows that localization is not possible under very general assumptions of quantum mechanics and special relativity. Further developments and associated theorem cement the case against localization.[9,10] Hegerfeldt's theorem is simple. In a one-particle theory, he starts by defining an operator $\hat{N}_V$ whose expectation value in a given quantum state $|\psi\rangle$, i.e. $\langle\psi|\hat{N}_V|\psi\rangle$, gives the number of particles in a volume *V*. He only assumes that the states that transform under the 10 irreducible representations of the Poincare group (4 translations, 3 rotation, 3 boosts) with positive energy and have positive or zero mass.[11] Then he shows, starting with a state, $|\psi\rangle$, that has probability *1* of having particle in finite volume *V*, that after that state is evolved for a time *t and* been spatially translated by an arbitrarily large distance to $|\psi'\rangle$, then it is mathematically impossible to preclude the particle from still being inside *V* (i.e., we cannot by any translation have $\langle\psi'|\hat{N}_V|\psi'\rangle=0$) ; i.e. the state has developed arbitrarily large tails. This shows that one cannot have a wavefunction that is localized as collapse would indicate in, for example, a two slit experiment.

---

[5] Such problems might still be posed as ways to probe the existence of theories beyond quantum mechanics, but in the mean time the meaning of current QM should be made as simple and clear as possible.

[6] Einstein asked A. Pais: "whether [he] really believed that the moon exists only when [he] look[s] at it.", A. Pais, Rev. Mod. Phys. 51, 863, p907 (1979).

[7] In a famous paper Wigner gives us the full blown consciousness collapses the wavefunction argument. E. P. Wigner, *Remarks on the Mind-Body Question*. In I. J. Good, (ed.) *The Scientist Speculates* (London: William Heinemann, Ltd., 1961; New York: Basic Books, Inc., 1962), ch. 13, pg 179. De Witt gives his classic defense of Everett's Many Worlds Interpretation: B. S. DeWitt, *Quantum Mechanics and Reality*, Phys. Today 23 No. 9, 30-35 (1970). R. Penrose, in *Road to Reality* (fn 18) (where he invokes gravity as the *R* process mechanism), and G.C. Ghirardi, P. Pearle, A. Rimini, *Markov processes in Hilbert space and continuous spontaneous localization of systems of identical particles,* Phys. Rev. A 42, 78 (1990), make valiant attempts to push beyond ordinary QM (e.g., beyond the SE). One must distinguish between such activity and the understanding of and application of quantum theory as it currently exists. Demonstrating that collapse is not only not necessary but can be seriously problematic takes away the original core reason for going this direction in seeking further theoretical advances, but it doesn't rule it out; after all they are, by definition, still seeking a form in which to be defined and then validated.

[8] G. C. Hegerfeldt, *Remark on causality and particle localization*, Phys. Rev. D 10, p3320 (1974).

[9] G. C. Hegerfeldt, *Violation of Causality in Relativistic Quantum Theory?*, Phys. Rev. Lett. 54 No. 3, 2395-2398 (1985) .

[10] G. C. Hegerfeldt, *Causality*, *Particle Localization and Positivity of the Energy*, Lecture Notes in Physics 504 (1998).

[11] He shows, in reference (fn 10), one only needs bounded from below Hamiltonian to develop infinite tails from localized particle localized in a finite region.

The two slit experiment will be our guide in exploring localization, collapse and particle/wave(field) paradox. In this article, we take a spiral approach reviewing as necessary while diving deeper into the understanding of collapse, localization and measurement in the two slit experiment. In addition, after building to it in the article, in the conclusion, we will answer: "are there particles or fields or what?"

In section IIa, we introduce the two-slit experiment as described in the standard (collapse) approach. In section IIb, the output of section IIa induces us to list four rudimentary facts about QM to keep in mind in our analysis and which will help expose the collapse postulate's failure. In section IIc, we return to explaining the two-slit experiment using the standard approach showing how, in that approach, the two-slit with detector (because it gives which way it went information) gives no interference build up over time, while the no-detector experiment yields interference. We expose the leaping over logic of the treatment of the wavefunction as representing a wave/particle that spontaneously collapses upon measurement to a wave in both slits or a particle in one. In section III, we use our analysis of the two slit experiment to now enumerate and examine four essential points about the nature of quantum mechanics, which build on the rudimentary points.

In section IV, we address the two slit experiment in the natural interpretation. In so doing we bring out the formalism and imagery to properly express the states. This includes giving the formalism to express collapse especially in its proper place as an approximation, not a postulate of the theory about how the state evolves. The collapse approximation is discussed simply using the state matrix theory for both the cases with detector and without. In section IVb, we directly address localization and the collapse approximation and their meanings. In section V, the importance of a proper definition of measurement is discussed and answered. Finally, in section VI, a clear example where the collapse postulate fails is given. The example shows how supposedly collapsed (i.e. completely gone) parts of the wavefunction actually interfere to radically change the predicted outcome.

For use throughout this article, we briefly introduce the standard (collapse) interpretation and the natural (ensemble) interpretation. While remaining sufficiently self-contained, this article is meant to further develop what has already been given in other works.[12,13,14] These other works have shown the importance of the natural interpretation and how little it is understood or even known.

The collapse interpretation is taught either explicitly[15] or implicitly to nearly all physics students. It is directly asserted as a postulate in many books.[16] The name

---

[12] A. Rizzi, *How the Natural Interpretation of QM Avoids the Recent No-Go Theorem*, Found. Phys. 50 No. 3, 204-215 (2020), A. Rizzi, *A Simple Approach to Measurement in Quantum Mechanics*, arXiv: Quantum Physics (2020), A. Rizzi, *Does the PBR Theorem Rule out a Statistical Understanding of QM?*, Found. Phys. 48 No. 12, 1770-1793 (2018) as well as in the new textbook referenced in (fn 13).

[13] A. Rizzi, *Physics for Realists: Quantum Mechanics* (PFR:QM) (IAP Press, Baton Rouge, 2018).

[14] See L. Ballentine's paper (fn 20) and textbook (fn 19).

[15] As an example take Shankar's book, which is popular (used by MIT, for example; see https://ocw.mit.edu/courses/8-04-quantum-physics-i-spring-2013)). R. Shankar: "when a system initially in $|\psi\rangle$ is measured, it changes to [i.e. collapses to] $|w\rangle$ (i.e., the eigenstate corresponding to measured value w)". R. Shankar, *Principles of Quantum Mechanics,* 2nd ed. (Plenum Press, 1994), pg 116. Another example is the popular book: D. H. McIntyre, *Quantum Mechanics: A Paradigms Approach*, (Pearson Addison-Wesley, 2012).

"Copenhagen" arises from the fact that it seems to derive from Heisenberg and Bohr, both working in Copenhagen, Denmark, but the first clear exposition seems to be due to Dirac and Von Neumann.[17] In the developed, "Copenhagen" view a measurement obeys a different law of evolution than the ordinary Schrödinger equation (SE). Roger Penrose uses the terminology: "*R* processes" and SE processes.[18] In an *R* process, the wave function jumps to an eigenstate of the operator corresponding to the observable property being measured. For example, a state described by $|\psi\rangle$ before a position measurement will, after the measurement, evolve through an *R* type evolution to $|x\rangle$, an eigenstate of position. Then, after the measurement is complete SE evolution takes over once again.

Now, the minimalist interpretation, called the ensemble interpretation, is a simple, natural approach to understanding QM. It is a newer approach first championed generically by Einstein but only developed formally by Ballentine,[19] starting with his seminal paper in 1970.[20] This approach was then, recently, more specifically and fully developed in a new textbook.[21] We summarize it as follows:[22]

The pre-measured state of the object to be measured can be represented by:

(1) $$|\psi\rangle^{(o)} = \sum_r c_r |r\rangle^{(o)}$$

The apparatus is represented by $|0,m\rangle^{(M)}$, where the readout is initially set at $\alpha = 0$

The resulting state after measurement, using only SE (unitary) evolution, is written:

(2) $$U|\psi\rangle^{(o)} \otimes |0,m\rangle^{(M)} = \sum_r c_r |\alpha_r;(r,m)\rangle$$

Where the superscript "*M*" indicates "measurement device."

And, the ket, $|\alpha_r;(r,m)\rangle \equiv \sum_{r',m'} u_{r,m}{}^{r'm'} |r'\rangle^{(o)} \otimes |\alpha_r, m'\rangle^{(M)}$ is a helpful notation that labels the complicated sum with the key variables: the initial state of the object, *r*, the initial *m* eigenstate of the measurement apparatus, as well as the resulting pointer (readout) value, $\alpha_r$.

In summary, the formalism captures the essence of measurement: an interaction that establishes a one-to-one correlation of the value of an observable property (corresponding to the eigenvalue *r*), of an object being measured at a particular point and an readout ($\alpha_r$). If the object is initially in the eigenstate associated with eigenvalue *r*, then the readout gives $\alpha_r$. Given the state in equation (1), there is a $|c_r|^2$ probability of getting $r$, therefore, as seen in equation (2) (and as it must be if the proper correlation exists), one gets the same probability of getting a reading $\alpha_r$ on the apparatus.

---

[16] More full exposition of how it is disseminated is given in the "Simple Approach to Measurement" and "Natural Interpretation" articles in (fn 12).

[17] Dirac discussed the jumping to an eigenstate after measurement in P.A.M. Dirac, *The Principles of Quantum Mechanics* (Clarendon, Oxford, 1930). Von Neumann made it clear in J. von Neumann, *Mathematische Grundlagen der Quantenmechanik* (Springer, Berlin, 1932). In English: J. von Neumann, R. Beyer (translator), *Mathematical Foundations of Quantum Mechanics*, (Princeton University Press, Princeton, 1955).

[18] Penrose called the latter "*U* processes," i.e. processes governed by unitary evolution. See R. Penrose, *Road to Reality: A Complete Guide to the Laws of the Universe*, (Knopf , NY, 2004), pg 528.

[19] L. E. Ballentine, *Quantum Mechanics: A Modern Development* (World Scientific Publishing, Singapore, 1998).

[20] L. E. Ballentine, *The Statistical Interpretation of Quantum Mechanics*, Rev. Mod. Phys. **42** No. 4, 358-381 (1970).

[21] See PFR-QM (fn 13).

[22] See (fn 19) and PFR-QM (fn 13).

## II: Two-slit Experiment: Standard (Orthodox) Approach

### IIa: Introduction

Consider the two slit experiment, which captures the core issues of quantum mechanics. We first analyze the particle passing through one slit. A very low intensity neutral atomic beam is pointed at two slits in a solid screen as shown in Figure 1. Because of the low intensity, one atom at a time emerges from the source. The single atomic wavefunction is shown as a Gaussian[23] in both spatial directions and traveling to the right centered on $x=0$; call this function $G_0$.

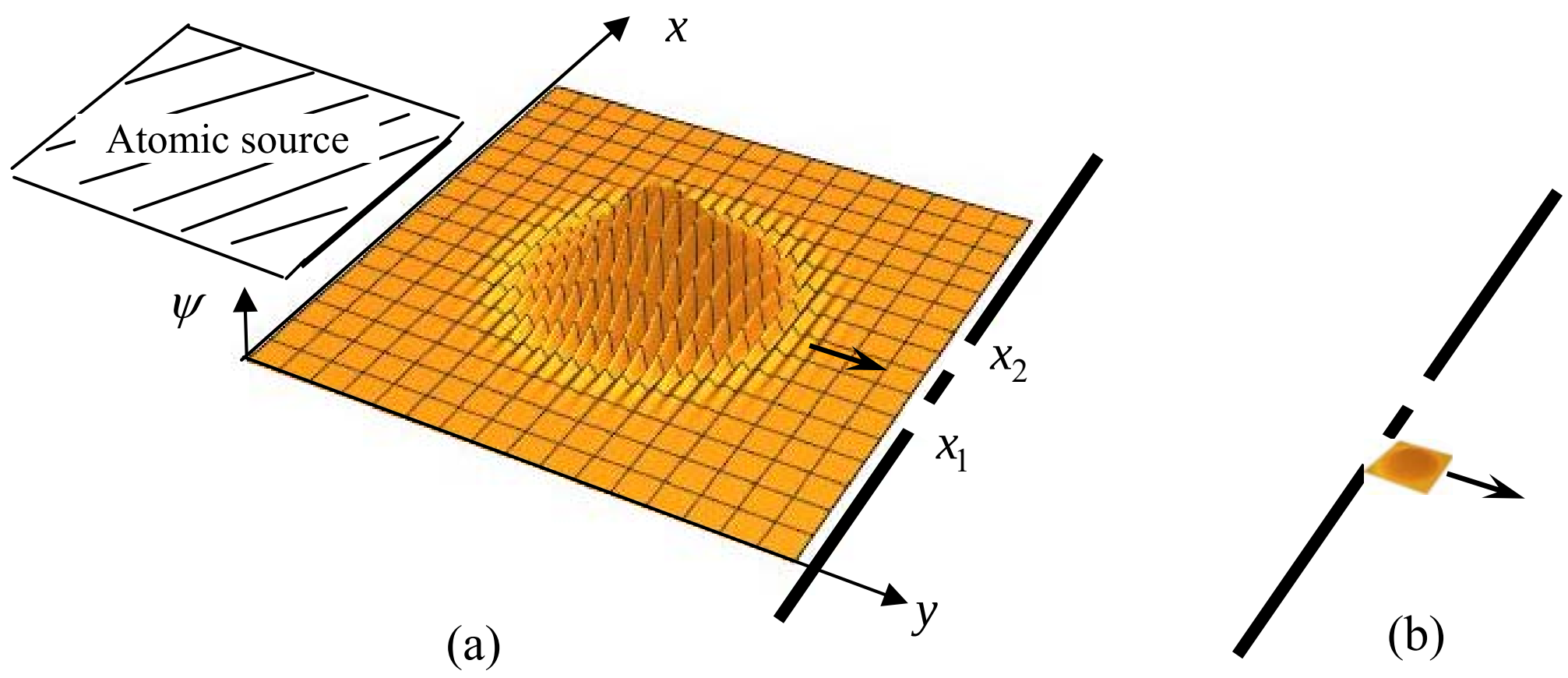

**Figure 1:** Schematic view of two slit experiment (detectors are inside of slits-- not illustrated): a) Gaussian electron packet impinging on two slits. b) After being detected at slit one, showing collapse of wavefunction centering on slit 1.

For symmetry in our current exposition,[24] there is a detector at each slit. For concreteness, we take the detector to consist of light beams on each side of each slit aimed at the slit's center so that if an atom passes through the slit, the atom is put into an excited state which then spontaneously decays by emitting light[25] that is detected at enough distance from the slit that the detection of that light doesn't interfere with the atom's further travel beyond the slits—label on figure $x_1$ and $x_2$ for each slit). If I measure the particle at the $x_1$ slit then, according the standard view, that measurement act collapses the particle wavefunction to the position eigenstate $\left|x_1\right\rangle$, as illustrated in Figure 1. We might refine this statement further by only requiring that that the atom be within a finite radius $\delta r$ from $x_1$ after time $\delta t = \delta r / c$. In other words, the system wavefunction, $\psi(x)$ changes ("collapses") from $G_0$ to something spatially confined to a finite region. Note this is not the evolution that the SE would predict. This is the so-called *R*-process already mentioned. After this point, however (in the standard interpretation), if no further measurement is made, we can evolve the wavefunction according to the SE. Before continuing this two slit experiment analysis in the standard approach, we need to point out serious problems with the interpretation.

---

[23] In the $y$ direction, the form that solves the free SE is: $\psi(y,t)=e^{-(y-\mathrm{v}t)^2/4\sigma_t^2}e^{i(ky-\omega t)}/((2\pi)^{1/4}\sigma_t/\sqrt{\sigma_0})$ where: $\sigma_t=\sigma_0\sqrt{1+it/\tau_0}$ , speed $=\mathrm{v}=\hbar k/m$ , and $\omega=\hbar k^2/2m$ .

[24] To reach the standard conclusions about quantum mechanics, one only needs a detector at one slit, but we want to start by analyzing the single slit case, so our setup is more elaborate than usually needed and typically employed.

[25] The atom is heavy enough and the light low enough energy that its path deflection due to emission is assumed extremely small.

What's wrong?

Most evidently, we have asserted an *R* process for no explicit reason and with no explanation as to why it is possible (we will shortly give an explicit reason and then show why it is not possible to insert an *R* process and still validate everything we currently know, which is captured in the SE); we only need SE evolution of the state. Another red flag, we have a localized wavefunction and, as mentioned, Hegerfeldt has shown that such a wavefunction will immediately develop infinite tails. This means that, in a moment, there will be a finite probability that the particle will appear arbitrarily far away! More precisely that there will be a detection indicating activity at localized detector arbitrarily far away, from which, therefore, we conclude the presence of a particle there. This, finally, means infinitely fast information transfer which is a violation of special relativity! This is not a problem for ordinary QM, which is not special relativistic, but it is a problem for the reality of the situation, which as far as we know, forbids faster than light information transfer (i.e. a measureable change in state that propagates faster than light).

How do we understand these leaps of logic and the problems these leaps cause. First, we need to become conscious of the fact that an operational point of view can be very helpful, even necessary, when first probing and solving a problem or class of problems, but this does not justify leaps in logic or a complete ignoring of why it worked or what the procedure developed means. It needs to be said that it's not, finally, even pragmatic to be solely pragmatic, i.e. to ignore or make it a principle to not know what one is doing, just do it. It can lead to erroneous results and, a fortiori, to fatal misunderstandings such as reality doesn't exist till it's measured, which is quickly deduced from collapse via Schrodinger Cat![26] Indeed, in ordinary QM, we have apparently, by default, hewed to a "we seem to use the equation in a certain way and it seems to work nearly all the time, therefore we can interpret the theory in the way that blends most simply with that use".[27] This is inadvertently sloppy. It doesn't allow enough space for careful logic based on what we know.

In order sort this out, let's now 1) catalog what we know, 2) continue the two slit explication and then 3) list the issues that we have not brought front and center. Once done, we can move to the next section to show how and why collapse is simply a useful approximation. It is one we have not understood as such and therefore not evaluated carefully enough, leading, among other things, to contradictions with Hegerfeldt's theorem. Start with the key things we know.

[26] See "Simple Approach to Measurement" and "Does PBR Theorem Rule Out" given in (fn 12).

[27] This might be called positivism, but calling it something doesn't eliminate its gratuitous skipping over many parts of the analysis of the measurement.

**IIb: Rudimentary Facts of QM**

*First*, we know that the SE is the valid evolution for the wavefunction in the SE's proper domain, which, of course, can be extended via Pauli equation[28] and higher order approximation[29] by incorporating some relativistic effects; this includes spin which often plays a center role in measurement discussions (and which we will discuss later). *Second*, we know from experiment that we must invoke a *probabilistic* theory at some level. Namely, experimentally, two quantum systems, set up in as an equivalent way as one can, do not yield the same results; so, any theory must describe the system in terms of probabilities, not individual outcomes. *Third*, experiments reveal, and it is evident from the wavefunction (which of course, itself came out of insight from experiments) that there is a wavelike interference behavior intrinsic to quantum mechanical systems. *Fourth*, quantum position measurements (as well as many others[30]) occur in particle like way; a measurement occurs at a location, e.g., a dot on a screen. A light beam or a massive particle beam "lights" up some very localized region of the screen when the intensity of the source is low enough. Let's investigate how we don't fully respect these facts in our analysis and draw out four facts that arise from QM itself, which once understood, themselves draw out and underscore the naturalness of the ensemble interpretation.

**IIc- Two Slit Continued: Interference and Non-interference**

To see this arise, we first complete the full standard (collapse) analysis of the two slit experiment. If we also, after measuring which slit each atom goes through (note, one only needs one detector on to do this), look at the screen behind the slits and let the dots left by each atom accumulate till we have sufficient density of dots to discern any pattern, we will see there is none! However, if the measuring devices are turned off at both slits so that there is no knowledge of "what slit the particle went through," then one will see the intensity interference pattern of the standard two slit wave experiment gradually build up as each atom leaves its dot on the screen behind the slits. This obviously indicates wave-like interference behavior. Measuring which way the atom went seems to change the nature of the system. But, we never see something as both a wave and a particle, because its contradictory. Collapsing the wavefunction upon measurement prevents us from seeing the superposition of going through both slits and not going through both, between being seen as *both* a particle and a wave. Upon measurement, according to collapse, it becomes one or the other.[31,32]

---

[28] $(((\hat{\mathbf{p}}-q\mathbf{A})^2-q\hbar\sigma\cdot\mathbf{B})/2m+q\phi)|\psi\rangle=i\hbar\partial_t|\psi\rangle$

[29] E.g, Darwin terms, and mass corrections.

[30] Take, as another example, measuring momentum by time of flight of a particle between two locations. Note: measuring energy in photo electric effect experiment (see PFR-QM, (fn 13)) at low enough intensity means one sees single events which are in fact "point" interactions that could be measured, but in such an experiment one, in principle, only needs to measure the occurrence of the event of the electron striking the plate, not the exact location at which the electron interacted.

[31] The response to "what about before it's measured" ends implicitly and often explicitly in "what isn't measured isn't seen so doesn't matter, at least to physics." This is despite the fact that many things that aren't measured in physics matter, such as the content of EPR and Bell's theorem which have deeply impacted pure and applied physics, even in as mundane things as cyber security.

[32] This wave/particle duality that somehow the particle is also a wave till it is measured and the superposed state is broken has been filled out further by Wheeler's which way "delayed choice" experiments, where

So, we have implicitly assumed the wavefunction describes a single entity and, because of that, we are forced, for example in the two slit experiment, to decide between which entity it is.

Again, this leaping over logic[33] is caused by a focus on what seems to work best with the formalism at the expense of sufficient careful analysis of what we are actually doing. Moreover, what seems to work by the seat of the pants (especially if we aren't fully aware that we are working that way) doesn't always actually work. In a later section, we give an example of the clear failure of a collapse prediction. Ballentine has previously discussed its limits.[34] For further analysis and how it can lead to wrong experimental results see the references given here.[19,35,36]

Again, we will show that the formal collapse mechanism (i.e. the state collapses to the eigenstate of the observable corresponding to the value measured) is a helpful approximation in limited cases and as such has its use. For now, let's further investigate where this *purely* operational mode of thinking leads and codify some conclusions that breaking out of this mode brings to the fore.

## III: Summary of Four Unknown, Essential Points about QM

In this operational mode (a mode that, unfortunately, extends beyond our practical thinking into our reflective, serious thinking),[70,72] we also tend to force the wavefunction to correspond to the result of a measurement or to measurements on multiple systems. Again, both this attempt and the attempt to force the wavefunction to represent a single entity are a result of our too quickly jumping to something that seems like a quick, straightforward use of the equation. And, neither one is correct. There are actually four essential points brought out by the two slit analysis keeping in mind the key quantum facts, especially those mentioned above. None are clearly discussed, if at all, in the literature.[37] Ignoring any one leads to confusion and even wrong results. The first two are of more general applicability, and the last two are more directly related to the Hegerfeldt localization. They are summarized here:

---

similar analysis is used to show collapse leads one to measurements being changed by the past! See J.A. Wheeler, *The "Past" and the "Delayed-Choice" Double-Slit Experiment*, Mathematical Foundations of Quantum Theory, pp. 9-48 (1978). See also: M. O. Scully, B. G. Englert, H. Walter, *Quantum Optical Tests of Complementarity,* Nature **351**, 111-116 (1991).

[33] For example, there is no necessity to this assertion as we seem to imply.

[34] L.E. Ballentine, *Limitations of the projection postulate.* Found Phys **20**, 1329–1343 (1990). See also PFR-QM (fn13).

[35] Murray Daw's "Avery Grace" (AG) experiment gives an example experiment in which a particle is in the "down" state directly after being measured in the "up" state! (private communications, to be published) The collapse postulate, of course, claims the opposite. One can claim that the particle was collapsed to the up state and then evolved into the down state. However, the measurement isn't complete until after the "pointer" is stably in its proper position, at which point the particle is in the down state. In any case, Daw's example starkly indicates that, in some cases, it is very easy to predict wrong results by invoking collapse.

[36] The literature sometimes discusses "hidden variables" when discussing the existence of various entities of QM that may not be included explicitly in the formalism. We already know QM is probabilistic and therefore leaves some entities out, so it's not helpful in this regard to introduce such terminology. Moreover, this very terminology tends to conflate (as mentioned in (fn 7)) the activity of discovering the next equation beyond QM (or possibly even QFT) with the activity of understanding and properly using QM as we now have it. The term "hidden variables" tends to imply that there are parameters that are waiting to be incorporated in one's formalism that one has just not been able to.

[37] Confer my papers and PFR-QM (fn 12,13) for more on these issues.

1) *The wavefunction has three elements* it *describes*: 1) a wavelike element and 2) a particle-like element and 3) a stochastic element. It's square gives, for example $P_1(x)=\left|\langle x|\psi_1\rangle\right|^2=\left|\psi_1(x)\right|^2$, the probability (stochastic agent acting) of finding the particle at a given point, $x$ (particle like agent acting) but $\psi(x)$ can interfere with other wavefunctions, to give for example $\psi_T=\psi_1+\psi_2$ (wavelike agent acting), to create different probability distributions, e.g. $P_T(x)=\left|\langle x|\psi_T\rangle\right|^2=\left|\psi_T(x)\right|^2$. Clearly, the wavefunction encapsulate the behavior of three entities, not one.

   a. There is no reason intrinsic to the theory to create a contradiction by forcing the wavefunction to simultaneously represent something localized and something not localized! We do it out of convenience, which turns out to be not convenient at all, but a source of much consternation and confusion!

2) *The wavefunction cannot be identified with the result of any single measurement or any group of measurements* (it can only deal with the statistics of groups of measurements). The wavefunction only deals with the probability of an event occurring and being measured, not what any given measurement result will be. Hence, QM cannot predict the state of any single system or its evolution afterward (for example, where a given particle will be found or, once we do know where it is, how it will move), for QM has from the outset left aside what happens to single systems. Instead, like a coin toss one is actually dealing with an ensemble of similarly prepared systems. In a coin toss we set each case up very like the one before but don't get the same result. In this way, we generate an ensemble of systems.[38] It's the same in QM, except that it is a matter of principle in the domain proper to QM. In the coin toss case, one can imagine creating an intricate machine that will toss the coin in a vacuum in a way that allows one to predict heads or tails reliably. Finally, of course, there will be a limit to the certitude which we can reach, for QM applies to coins as well as atoms; however, note that we are drawing the distinction between quantum level and classical level physics which, by definition, leaves those QM behaviors behind. So, now we are ready to see our second issue in its particular form:

   a. According to the collapse postulate, every measurement (in particular the new knowledge or potential knowledge of the system gained by the (potential) measurement) forces us to evolve the wavefunction outside of is normal SE development.[39] In our example, finding the particle at the first slit, forces us to change by my act of will from $|\psi\rangle$ to $|x_1\rangle$. However, as just stated, the formalism knows nothing about single events, only the statistics of an ensemble of events! As much as we may want it to, QM cannot help us follow the evolution of a single system! It only works on ensembles of systems. Now, in taking, for example, only those systems in which we find the particle at slit one, we are *not* thereby reducing the description of those systems to simply "atom is here" and that is all there is to say about the systems." Still, less can we reduce it to the formal state $|x_1\rangle$.[40] Instead, we are creating a subensemble of the larger ensemble (the larger being the one where we make no distinction about where the particle will be found). It is this subensemble whose subsequent statistical behavior we can then describe using the SE equation (if done with proper care using the known wavefunction of the full ensemble). By measuring the particle, we *do*, indeed, change the state, for the measurement invokes a new potential that changes the SE for a while.[41] Of course, this does not mean we don't also know something about the single system.

[38] Of course, initially we just have the possibility of doing such an experiment and thus only a potential ensemble. The experimenter creates the ensemble and then does the experiments (both in the case of the coin and the quantum system).

It means that what we know after a measurement cannot be substituted for the quantum state specified by the wavefunction. Instead, it's as if in the act of flipping the coin and finding it heads, we change something about the coin flip system so that it is no longer ½ chance of being heads but only ¼; we know it was heads but we don't know it will stay heads next flip, we only know its probability.

Recapitulating more detail, after collecting the sub-ensemble of systems, we know one particular thing about each system in the ensemble: we know the atoms' position in some fuzzy way at some time. However, we don't know exactly how the single particle changes during and after the measurement, neither do we, in general, know its momentum well enough (even if there were no stochastic agents) to make very helpful predictions of the atom's future motion. Furthermore, after the measurement, we still neither know the singular state (or its evolution) of the guiding wave-structure nor the singular state of the stochastic agents (or its evolution) for any given experiment. Again, as before the measurement, the only tool we have to tell us something about the nature of the sub-ensemble that has the atom near position *x* at time *t* is the wave-function of the full ensemble and how it evolves according to the SE and that contains statistical information only, not information about any given system.

[39] In my experience, many tend to think (perhaps by thinking of spin or energy eigenstates) that the collapse postulate means that the measurement changes the state to an eigenstate and then is frozen in that eigenstate until further interaction. However, this doesn't make any sense. To make the collapse postulate make sense, we need to say (as we've alluded to and done in text, e.g. p5) that after collapse by measurement, the state continues again to evolve according to the SE. It's part of the ad hoc nature of the collapse postulate to be only vaguely pragmatically defined and thus fall prey to these kind of issues.

[40] Even less is our *knowledge* of the measurement result (which is necessary in order to make any decisions about the to classify the systems) causing such a reduction of $|\psi\rangle$ to $|x_1\rangle$.

[41] But, again, let me underscore, the act of measurement doesn't make us move from a statistical theory to a theory of individual particles.

In summary, *we must distinguish between the quantum state, e.g.* $|\psi\rangle$ or $\psi(x)$, *and the physical state of system, i.e. the actual state of each of the physical elements in a given individual system. We are in the habit of spontaneously identifying the two as evidenced by our use of the word "state" without such qualifiers.* Quantum mechanics is not like classical mechanics, where, for example, *x(t),* signifies the position of a single particle at given time in a given system, which then evolves under Newton's second law.

*3)* *The wavefunction does not fully describe known reality, even probabilistically.* Quantum field theory makes this evident.

a. In particular, one cannot pick just any wavefunction, as it might violate other physics such as special relativity (SR). The key example is the one discussed earlier in the introduction to the two slit problem in which one tries to localize a wavefunction and runs afoul of Hegerfeldt's theorem.

*4)* *One must distinguish between approximation and the theory itself.*

Seeing this distinction and making explicit and clear the nature of the approximation that is misnamed "collapse" brings forward the key mechanism by which the introduction of collapse as well as the tendency to force finite localization came to be self-sustaining problems in understanding quantum mechanics.

## IV: Two Slit Experiment Properly Understood

Finally, to fully understand the problems around collapse and localization, we continue our spiral technique by looking at the two slit experiment again with this knowledge in hand. In so doing, we will bring out and apply the natural (ensemble) interpretation. To do this, we lay out the requisite formalism.

We will make implicit use of the many-particle Schrödinger Equation, which, as given, applies to non-relativistic systems, which, in contrast to full blown QFT, therefore has a fixed number of particles.

$$\hat{H}\Psi(\vec{x}_1,\vec{x}_2...\vec{x}_n)=i\hbar\dot{\Psi}(\vec{x}_1,\vec{x}_2...\vec{x}_n) \text{ with } \hat{H}=\sum_{i=1}^{n}\frac{1}{2m_i}\vec{p}_i^{\,2}+V(\vec{x}_1,\vec{x}_2,\vec{x}_3...\vec{x}_n)$$

where $\vec{x}_i$ is the position of the $i^{\text{th}}$ particle.

From this, it is clear that, in general, there is one total wavefunction that is a function of all configuration space, which means the probability of finding one particle is also dependent on the location of the other particles. This is a kind of complicated quantum entanglement. In our analysis, we assume that the detectors, screen and electron are initially isolated from each other sufficiently so that the total wavefunction can be written as a product: $\psi=\psi_{\det}\psi_{elec}\psi_{screen}$, thereby assuming effectively no entanglement.[42]

---

[42] In the early universe everything was interacting, but we assume over time, the interactions were such that as their hold loosened on each other, this analysis became effectively valid.

To both write the explicit wavefunction for the system at various times and from there to give the helpful state matrix formulation, we draw a schematic outline of the two-slit experiment in Figure 2 and proceed to analyze the with-detector and without-detector cases.

### IVa: Deriving Quantum States and Introduction State Matrix formalism With-Detector Case

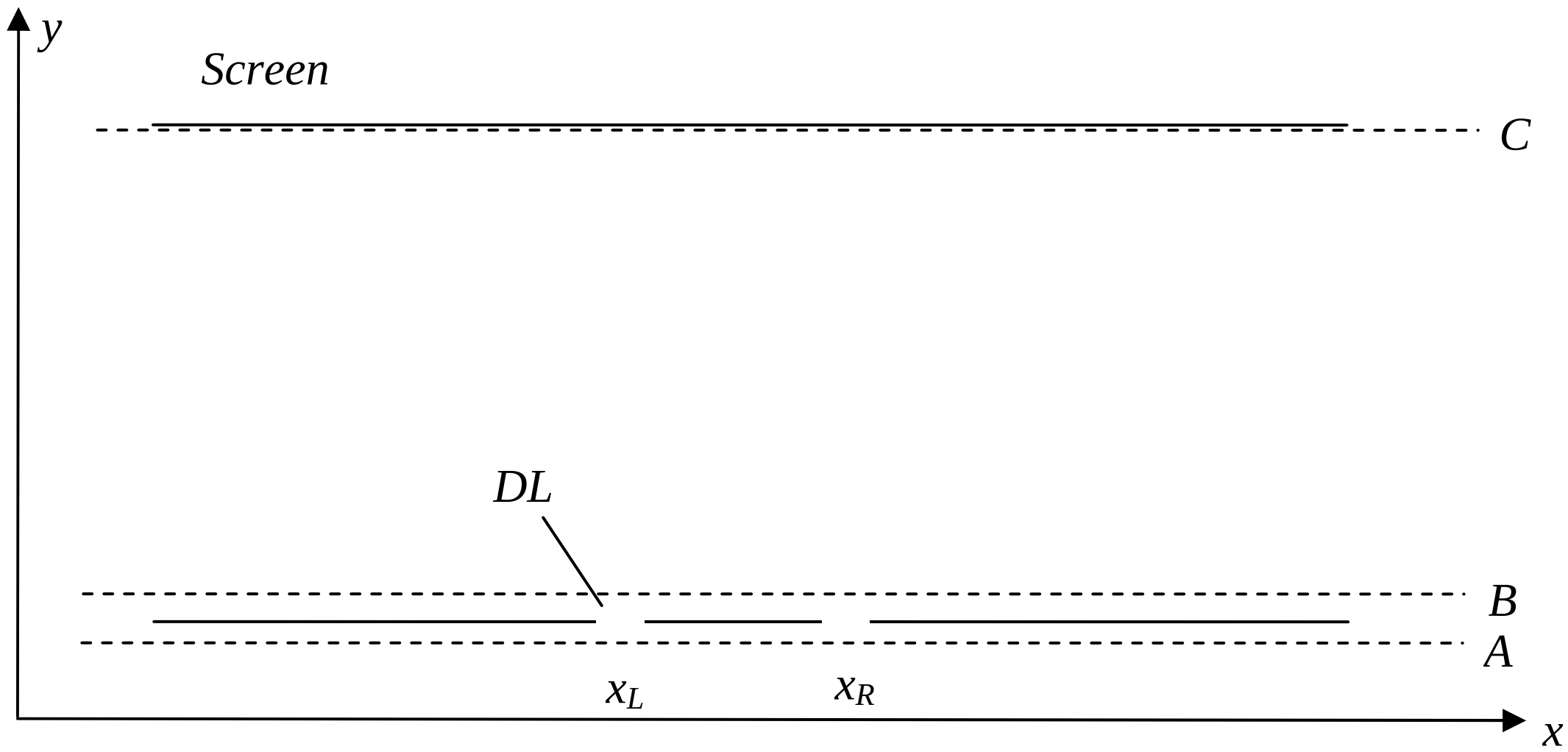

**Figure 2:** Schematic Outline of 2-Slit Experiment

Before impinging on the slits shown in the figure, the incoming wave (not shown) can be described by a traveling wavepacket. As discussed and shown in Figure 1, the wavepacket can be taken to be effectively a plane wave, $e^{i(ky-\omega t)}$, of frequency $\omega = E/\hbar$, where $E$ is energy of each atom in the atomic beam, with a Gaussian envelope in *x* and *y* (see Figure 1).
Then, right before wavepacket entrance (slice A in figure), one can write the total state as:

$$(3) \qquad |\Psi_A\rangle = c\left(\frac{1}{\sqrt{2}}\left(|\psi_{xL}\rangle + |\psi_{xR}\rangle\right) + c_e|\psi_{else}\rangle\right)|DL_0\rangle\,|S\rangle \quad (\text{at } t=0, y=0)$$

There are three objects considered: the atom, the detector, which is at the left slit (designated *DL*, with subscript *0* indicating detector in non-detection state) and the screen (S). As already stated, they are initially taken to be untangled; i.e., in the single particle approximation of ordinary QM, we take them to be described by three kets multiplied by each other, each evolved according to its own separate Schrödinger equation as long as there is no interaction between the three components of the system.

We have here taken the atom state description to consist of three parts, the part of the atom wavefunction at the left slit, the right slit and everything else. In particular, $|\psi_{xL}\rangle$ is part of the wavefunction that impinging on the left slit[43] and $|\psi_{xR}\rangle$ is part of the wavefunction that impinging on the right slit and the remainder is dubbed "everything else". Note that in the actual physical world, the screen will never totally confine the atom; i.e., some atoms, though extremely low probability, will penetrate the screen in places where there is no slit (so the wavefunction will have tails!).

[43] This means it is confined spatially to a finite region, which would be evident in the form of $\psi_{xL}(x) \equiv \langle x|\psi_{xL}\rangle$).

Now, we take the detection to be confined spatially to the left slit, so that its spatial wavefunctions domain is almost exclusively within the left slit.
After moving through the slit, we now have, at slice B in Figure 2:

(4) $$\left|\Psi_{\mathbf{B}}\right\rangle = c\left(\frac{1}{\sqrt{2}}\left|\psi'_{xL}\right\rangle\left|DL_1\right\rangle + \frac{1}{\sqrt{2}}\left|\psi'_{xR}\right\rangle\left|DL_0\right\rangle + c_e\left|\psi'_{else}\right\rangle\left|DL_0\right\rangle\right)\left|S\right\rangle \quad (\text{at } t=t_B, y=y_B)$$

The prime on the atom wave-states distinguishes the after-going-through-slitted-screen wave-state from the one before. Here we also see that the atom's interaction with the detector changes the state of the detector depending on which slit the atom actually goes through, and it, thereby, leaves the detector entangled with two parts of the atom wave function: 1) "the left slit" and 2) "the right slit + everything else."

With the particular equational structure and its proper context in place, we now describe its use. Equation 2 shows that the probability of finding the atom in slit one is:

(5) $$\left|\left\langle\psi'_{xL}\right|\left\langle DL_1\right|\left|\Psi_{\mathbf{B}}\right\rangle\right|^2 = \frac{c^2}{2} \approx \frac{1}{2}, \text{ (because } \left|c_e\right|^2 << 1 \text{ so}^{44} \left|c\right|^2 \approx 1\text{)}$$

But, of course, one's act of finding this probability does not reduce the state $\left|\psi_B\right\rangle$ to just the first term. Neither, of course, therefore, does one's use of this measurement to pick out a subset of those systems (i.e. those that have clicked detectors) change the overall system's generic nature. Of course, I do know (now) that this subset of systems, all have their atom at left slit, but that information itself doesn't change the further evolution of the full ensemble, which is described only by the SE. One can make an approximation in which we ignore all the members except those that don't have their atom at slit one, but it is just that, an approximation of some kind and what we do beyond selection has to be done with care. Again, the system is described by eq (4) (and its SE evolution) *whether or not* I choose to focus on only a labeled subset of the full ensemble of systems. The only physical changes my two mental actions cause[45] occur within me; in particular, in my neurological, especially brain, structure change. These changes, if the experiment is done properly, have minimal affect on the relevant (two-slit) part of the system. Notice there is a crucial difference, however, between the with-detector experimental setup just described and the without-detector experimental setup. (Keep in mind that each experimental setup, to be useful in ordinary QM, must be run many times. This means we always associate each experimental setup with an ensemble of physical systems).[46]

---

[44] This derives by applying normalization to $\left|\Psi_A\right\rangle$ and getting $\left|c\right|^2(1+\left|c_e\right|^2)=1$.

[45] I.e., 1. my action of calculating the probability and 2. my focusing on those members of the ensemble that have the atom at the left slit.

[46] Each physical system (though having much in common due to the fact that each corresponds to the same experimental setup to the degree we, in principle, can externally make them so) does have a different 1) guiding wave structure, 2) atomic structure and 3) unidentified stochastic agents. Note a distinction that may be helpful: if the experiments are run, we refer to actual systems; if they not, we speak of an ensemble of potential systems (that would arise from experiments if we did them).

### Without-Detector Case

We saw at slice B shown in Figure 2, the with-detector setup was described by quantum state $|\psi_B\rangle$ given in equation 2. By contrast, without the detector the system state is:

$$(6)\qquad |\overline{\Psi}_{\mathbf{B}}\rangle = \overline{c}\left(\frac{1}{\sqrt{2}}\left(|\psi'_{xL}\rangle + |\psi'_{xR}\rangle\right) + \overline{c}_e\,|\psi'_{else}\rangle\right)|S\rangle$$

Here we see that there will be visible interference between the two slits at the screen, if we note that $\overline{c}_e << 1$ is very small. Think of the two slits as producing coherent, in-phase single-particle wavefunctions that interfere, similar to the way the approximately spherical waves emanating from the two slits in optical experiment would.

This is very unlike the system given by equation 2, for in the with-detector setup, the waves can no longer interfere because the waves from the different slits are of a different nature and can no longer interact in the simple way that they could before the detector. This is caused by the presence of different detector wave structure in slit 1. This is represented formally by the left slit state being multiplied by $|DL_1\rangle$ and the right by $|DL_0\rangle$. Now, this is very like the classical, optical version of the two-slit quantum eraser experiment outlined in PFR-QM.[47] In that case, the different polarization light at each slit (one vertical one, horizontal) makes it so interference cannot occur, even though the *E* and *B* wave structure (or simply *A*-field)[48] are still present. This polarization difference is analogically like the differing guiding wave-structures emanating from each of the two slits due to the simultaneous action of the detector and atom wave-structures. In both this with-detector case and the classical optical case, one sees a Gaussian blur on the screen instead of the standard two-slit interference structure.

### State Matrix: with and without detector Cases

Now, we are ready to review the standard state matrix notation to help formalize the implementation of the "collapse" approximation.

Dropping the screen state ket, the state matrix of ***with detector case*** is:

$$(7)\qquad \rho = |\Psi_B\rangle\langle\Psi_B| = |c|^2\left(\frac{1}{\sqrt{2}}\left(|\psi'_{xL}\rangle|DL_1\rangle + |\psi'_{xR}\rangle|DL_0\rangle\right) + c_e\,|\psi'_{else}\rangle|DL_0\rangle\right)\cdot$$
$$\left(\frac{1}{\sqrt{2}}\left(\langle\psi'_{xL}|\langle DL_1| + \langle\psi'_{xR}|\langle DL_0|\right) + c^*{}_e\,\langle\psi'_{else}|\langle DL_0|\right)$$

In this basis $\rho_{mn} = \langle\phi_m|\Psi_B\rangle\langle\Psi_B|\phi_n\rangle$, where $|\phi_k\rangle = |\psi_k\rangle|D_k\rangle$,[49] and we can write this as:

[47] See PFR-QM (fn 13), chapter 8 problem 34, page 473.
[48] A. Rizzi, *Physics for Realists: Electricity and Magnetism* (IAP Press, Baton Rouge, 2011).
[49] Note: $|D_1\rangle = |LD_1\rangle, |D_2\rangle = |D_3\rangle = |LD_0\rangle$

(8)

$$\rho=|c|^2 \overbrace{\begin{pmatrix} \frac{1}{2} & \frac{1}{2} & \frac{c_e}{\sqrt{2}} \\ \frac{1}{2} & \frac{1}{2} & \frac{c_e}{\sqrt{2}} \\ \frac{c^*_e}{\sqrt{2}} & \frac{c^*_e}{\sqrt{2}} & |c_e|^2 \end{pmatrix}}^{|\psi'_{xL}\rangle|DL_1\rangle \quad |\psi'_{xR}\rangle|DL_0\rangle \quad |\psi'_{else}\rangle|DL_0\rangle} \left.\begin{matrix} \langle\psi'_{xL}|\langle DL_1| \\ \langle\psi'_{xR}|\langle DL_0| \\ \langle\psi'_{else}|\langle DL_0| \end{matrix}\right\}$$

Notice the presence of the off-diagonal elements which can, in the right cases, signal the presence of interference, which happens in pure states, not mixed states (which are necessarily purely diagonal). However, they do not imply interference in this case, for, as was already said, there is none. We need to implement the collapse approximation to make this evident, which we do by a process called reduction. For the moment, for comparison, consider the without-detector case.

The state matrix in the *without detector case* is:

(9)

$$\bar{\rho}=|\bar{\Psi}_B\rangle\langle\bar{\Psi}_B|=|c|^2\left(\frac{1}{\sqrt{2}}\left(|\psi'_{xL}\rangle+|\psi'_{xR}\rangle\right)+\bar{c}_e|\psi'_{else}\rangle\right)\cdot\left(\frac{1}{\sqrt{2}}\left(\langle\psi'_{xL}|+\langle\psi'_{xR}|\right)+\bar{c}^*_e\langle\psi'_{ekse}|\right)$$

$$\bar{\rho}=|c|^2 \overbrace{\begin{pmatrix} \frac{1}{2} & \frac{1}{2} & \frac{c_e}{\sqrt{2}} \\ \frac{1}{2} & \frac{1}{2} & \frac{c_e}{\sqrt{2}} \\ \frac{c^*_e}{\sqrt{2}} & \frac{c^*_e}{\sqrt{2}} & |c_e|^2 \end{pmatrix}}^{|\psi'_{xL}\rangle \quad |\psi'_{xR}\rangle \quad |\psi'_{else}\rangle} \left.\begin{matrix} \langle\psi'_{xL}| \\ \langle\psi'_{xR}| \\ \langle\psi'_{else}| \end{matrix}\right\}$$

This clearly has cross terms as well. However, there is a significant difference, in that these terms are not mixed with detector terms! This means, as discussed by analogy earlier,[50] these *can* interfere.

Now, we implement our approximation of ignoring the detector states. This will make evident the non-interference of the with-detector case. It reduces the state (in our mental approximation, not of course, physically).

In particular, in the *with-detector case*, tracing $\rho$ over the detector states (thereby ignoring them) gives the reduced matrix, using equation (6):

(10) $$\rho_{reduced}=\mathrm{tr}_{\det}\,\rho_{mn}=tr_{\det}\left(|\Psi_B\rangle\langle\Psi_B|\right)=\sum_{m=1}^{2}\langle D_m|\Psi_B\rangle\langle\Psi_B|D_m\rangle$$

$$=\langle DL_1|\Psi_B\rangle\langle\Psi_B|DL_1\rangle+\langle DL_0|\Psi_B\rangle\langle\Psi_B|DL_0\rangle$$

[50] For discussion of how this happens in more detail in a QM case, see PFR-QM (fn 13) chapter 8, problem 35 p475.

$$=|c|^2\left(\frac{1}{2}|\psi'_{xL}\rangle\langle\psi'_{xL}|+\frac{1}{2}|\psi'_{xR}\rangle\langle\psi'_{xR}|+|c_e|^2|\psi'_{else}\rangle\langle\psi'_{else}|+\frac{c_e}{\sqrt{2}}|\psi'_{else}\rangle\langle\psi'_{xR}|+\frac{c^*_e}{\sqrt{2}}|\psi'_{xR}\rangle\langle\psi'_{else}|\right)$$

$$\rho_{reduced}=|c|^2\begin{pmatrix}\frac{1}{2} & 0 & 0\\ 0 & \frac{1}{2} & \frac{c_e}{\sqrt{2}}\\ 0 & \frac{c^*_e}{\sqrt{2}} & |c_e|^2\end{pmatrix}$$

Now, in the two slit sector, we have only diagonal elements, which signals ***no interference*** and our formalism corresponds with our simple experiment.

**IVb: Localization and Collapse as an Approximation**

To fully explore the nature of the collapse approximation, we first eliminate the $\psi_{\text{else}}$ portion of the state given in (6). It may seem that this part of the wavefunction should never have been added as it is so radically unlikely for a particle to get through the screen anywhere but through the slits. However, deleting it gives:

(11) $$|\bar{\Psi}_{\mathbf{B}}\rangle=\bar{\bar{c}}\left(\frac{1}{\sqrt{2}}\left(|\psi'_{xL}\rangle+|\psi'_{xR}\rangle\right)\right)|S\rangle$$

and thereby confines the particle to only being in a finite part of space at a given time. This means, according to Hegerfeldt's analysis, that a moment later infinite tails will develop! Deleting the "$\psi_{else}$" is not as inconsequential as one might think! We should have known, for item (3) of the unknown essential points warns us the wavefunction does not describe all known reality so cannot be expected to made whatever we want to force it to give a result in correspondence with measurement.

In particular, as we already saw, this localization problem immediately appears when we try to argue (and professors teaching elementary courses often do) for working with position eigenstates in the formalism in the following way (and thereby effectively introduce collapse via an implicitly approved generic method: the practical route). We argue, as outlined earlier: "a particle is detected at the slit; it can go no faster than the speed of light so it must be localized near the slit, so $\psi$ is spatially confined to finite region." This, again, is not possible as infinite tails would instantly develop and make it possible to be anywhere.

But, how does this work? It cannot go faster than light, but it seems it must? No it doesn't go faster. Remember the only evolution of the wavefunction that we have is the SE.

Now, Barat and Kimball[51,52] have shown that neither the Klein-Gordon, nor the Dirac equation (and, therefore, not the SE non-relativistic limit) allow solutions whose tails fall faster than a certain rate. Therefore wavefunctions cannot be confined locally to a finite region.

[51] N. Barat, J.C. Kimball, *Localization and Causality for a Free Particle*, Physics Letters A 308 (2003) 110-115.

[52] This is argued by showing that the requisite exponential fall off for finding the particle in a volume as the volume is expanded is inconsistent with the analytical conditions required for solutions of the KG and Dirac equations.

This simply reminds us we cannot assert any wavefunction we want. Instead, we must remember that the wavefunction does not and cannot represent a single particle or even a single system; it represents an ensemble and not just any ensemble! So, we shouldn't try to force localization on the wavefunction; we must remember that there are parts of the system one has ignored that have particles in the tails and those tails matter in some cases, but they *always* matter when it comes to understanding what we are doing. Moral: 1. don't fall for the temptation to force the wavefunction to correspond directly to your single value measurement (or group of them) and 2. blind pragmatism is not pragmatic.

What do we do when we want to know about the subset of systems with the particle confined to the slit area? The subset doesn't make a proper ensemble for use with the wavefunction, but we have QFT, which (reducing to SE in the proper limit) tells us more about the system.

QFT shows that the $\psi$, is actually a quantum field, which contains information about the system beyond where the particle is probabilistically (and something about its guiding wave structure) in the non-relativistic limit. When one tries to force a wavefunction to be confined to less than a Compton wavelength, particle-antiparticle pairs begin to be created and particle number is therefore not conserved, so the field's non-relativistic use as a simple wavefunction fails. However, the field contains important information,[53] including information about the interactions within the guiding wave structure. Relevant here is the Reeh-Schlieder Theorem in QFT.[54,55] It implies that interactions can propagate faster than light from any point in space to an arbitrarily far away place. However, it can also be shown that unitary operations (i.e., "physical" ones, ones that transmit measurable interactions[56]) between space-like-separated points *cannot* exist.[57] In short, QFT shows that non-measureable superluminal interactions occur even in vacuum, but measurable interactions *cannot* occur between space-like separated points.

So, we can say the particles in the subset cannot traverse further than the speed of light from their initial detection place. However, we cannot force the wavefunction to be that without contradiction.

The problem is that we have made an approximation. Collapse is an approximation. We have assumed the wavefunction now is simply near the slit after we measure it. This is false. The image shown in figure 1 is false in multiple ways. The wavefunction doesn't collapse! Key is: the guiding wave structure continues to exist in all systems, including of course, the subset I pick by choosing ones found near the slit! Seriously erroneous predictions arise if one doesn't keep this in mind. A later section provides a clear example of this using a Mach-Zehnder interferometer.

---

[53] Most often captured by scattering analysis into non-interactive zones where particle number is constant.

[54] H. Reeh, S. Schlieder, *Bemerkungen zur unitäräquivalenz von lorentzinvarianten feldern*. Nuovo Cim **22**, 1051–1068 (1961). https://doi.org/10.1007/BF02787889 (apparently not available in English (title is: "Remarks on the unit equivalence of Lorentz invariant fields"). Cf: in English, E. Witten, *Invited article on entanglement properties of quantum field theory*, Rev. Mod. Phys. **90**, 045003 (2018).

[55] It says that an arbitrary state $|\psi\rangle$ in Hilbert space can be created from the vacuum (or any state with bounded energy) through use of operators whose support is purely local, even for states that have excitations arbitrarily far away. They can still generate quasi-local states but not properly local not states with support only locally; there will always be tails; still subluminal travel of a massive particle, in terms of detection at only time-like separated points is guaranteed by commutation (anti-commutation relations). This can be shown to be related to quantum entanglement in ordinary QM.

[56] Only unitary operators (acting on states) conserve probability which is essential for the theory to have predictive value or even a proper meaning.

[57] https://rojefferson.blog/2018/04/08/the-reeh-schlieder-theorem/

We need to make clear our "collapse" approximation then. We consider only a subset of the actual systems. We then approximate the wavefunction by a local function of finite support to capture the relativistic fact (we can model this local function as a Gaussian which we evolve according to SE, but which we finally take the limit to get arbitrarily close to a delta function). In so doing, we ignore the rest of the wavefunction that goes through the other slit, as well as the small amount in $\psi_{\text{else}}$; this means ignore the tails that are naturally in the actual wavefunction that we approximate.[58] So, we have culled the original system and then in each of those systems we have trimmed out some of the causal elements of the system. The approximation is good if the proper conditions are met and, thereby, explains the overall success of collapse as a tool. Still, of course, such success doesn't obviate the need to pay attention to what is being done.

So, having eliminated the $\psi_{\text{else}}$, we now proceed to implement the collapse approximation using the formalism, leaving us with equation (11), which gives, for the just discussed *with-detector* case, after tracing over the detector states:

(12)
$$\rho_{reduced} = |\bar{c}|^2 \begin{pmatrix} \frac{1}{2} & 0 \\ 0 & \frac{1}{2} \end{pmatrix}$$

We see that the cross terms are non-existent. This reflects the fact that we have destroyed any possibility of interference by our approximation that the wave only goes through one slit, (which, in turn, reflects the fact that the two parts of the wavefunction cannot interfere).

Meanwhile in the *without-detector* case, we keep both parts of the wavefunction and this shows up in the formalism by giving the proper cross terms for interference:

(13)
$$\bar{\rho} = \frac{|c|^2}{2} \begin{pmatrix} 1 & 1 \\ 1 & 1 \end{pmatrix}$$

Our formalism nicely implements collapse, but, let me emphasize it is an approximation where we have deliberately left parts of the formalism behind (and therefore the parts of the reality reflected in that formalism).

As one might expect given the prevalent misunderstanding of collapse, we should note that applying this state matrix formalism to the collapse postulate has led to claims that the formalism itself does things it cannot do. Ballentine has addressed one of these head on.[59]

## V: Measurement in Natural Interpretation

To fully digest the problem with collapse and to properly digest its subsequent treatment as merely an approximation, we must have a clear understanding of measurement. In particular, the question: “What is measurement?” must be clearly answered. As explained in the introduction in fully formal form, a measurement is an interaction after which results a one to one correlation between the measuring apparatus and the system to be measured. In the habit of using collapse, we can get stuck on the idea that measurement must leave the measured system in an eigenstate[60,61] and thereby make that (seemingly innocuous and clean) statement the definition of measurement. Let’s be clear, this is not at all required of a measurement. We just want to know what the value of the observable is at the point of measurement, no more.

---

[58] Obviously only certain types of measurements fall in the category that this approximation works. See (fn 19) and PFR-QM (fn 13) for discussion of filter measurements.

[59] L. E. Ballentine, *Classicality without Decoherence: A Reply to Schlosshauer*, Found. Phys. 38, 916-922 (2008).

Before closing, we give an example that shows the dangers of not recognizing collapse as an approximation in which certain things are left out for convenience of calculation.

## VI: Example where Collapse Fails to Give Correct Prediction

Consider a Mach-Zehnder interferometer in a tuned state (even after any lengthening devices are inserted). In our thought experiment, which is shown schematically in Figure 3, we inject atoms into the interferometer. The first “mirror” is a Stern Gerlach device, sending spin up atoms upward and spin down to the right.[62] All other mirrors are standard; the top two mirrors being fully reflective of the atomic beam and the last mirror (lower right) being partially reflective so that half the atomic beam is reflected to the right and half is transmitted downward. The interferometer is tuned such that the atoms can only come out of the “light port,” which is defined as the rightward output of the lower-right mirror (with the continuing line emerging), as opposed the downward output (with only a line fragment emerging). In short, it is tuned such that there is complete constructive (destructive) interference at the rightward (downward) output.

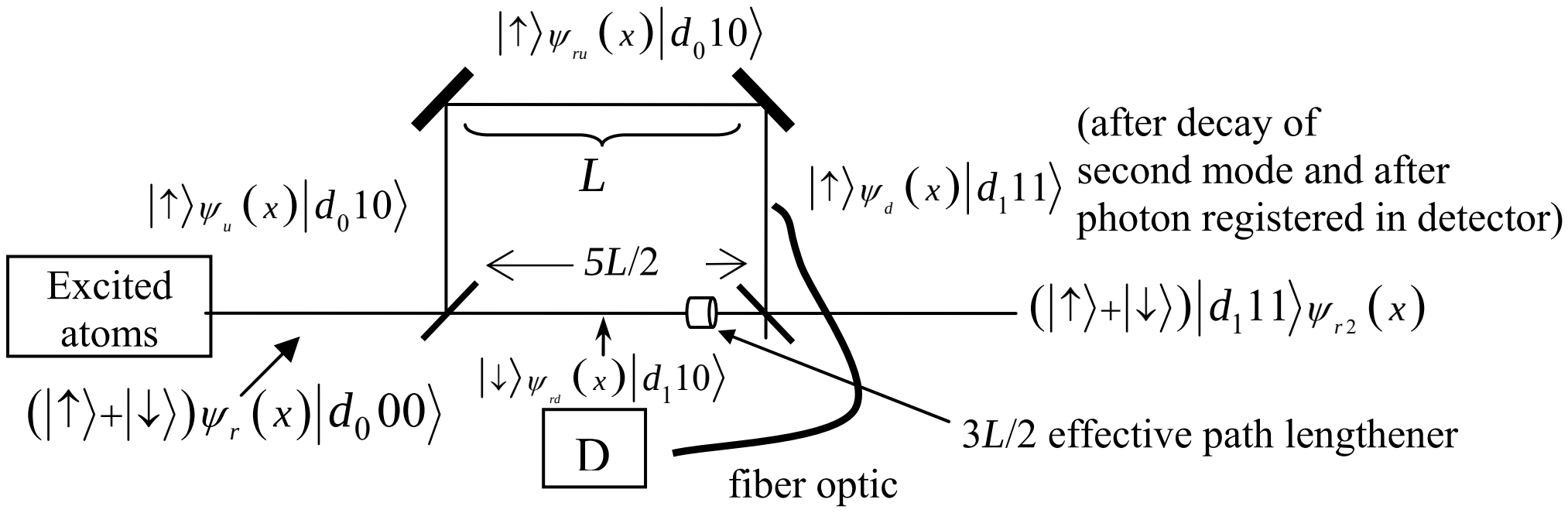

**Figure 3:** Mach-Zehnder interferometer which shows interference even after “collapse” thereby showing that assuming a “collapse” evolution (*R*-process) can give incorrect predictions.

---

60 The quantum Zeno paradox results from this mindset. Ballentine responds to it in reference given in (fn 61). It fundamentally results from the particular fields used in the experiment, not from collapse.

61 L. E. Ballentine, *Comment on “Quantum Zeno effect”*, Phys Rev. A 43 No. 9, 5165-5167 (1991).

62 Each atom starts with $l=2$ orbital angular momentum, which it loses before exiting the MZ by emitting two photons. We don't keep track of this side process, except to say at the first "mirror " the Stern-Gerlach splits the incoming beam into 6 beams, and we absorb (and thereby ignore) all except the + 5/ 2 and +3/2 beams. Hence, after emitting two photons, the top beam has total angular momentum 1/2 and bottom -1/2. The halves are due to spin angular momentum, which we do keep track of. Namely, halves refer solely to spin state, ignoring the orbital angular momentum

Each atom has two distinct modes excited; the first mode is timed to decay and emit a photon half way through first leg (at about $L/2$ distance traveled past first mirror), and second mode is timed to decay at half way through the third leg (at about $5L/2$ distance traveled past first mirror). Note the length extender located in the lower leg (the one defined by beam that goes straight through the input mirror on the lower left).

Each atom enters doubly excited and in a combination of spin up and spin down states (say spin right).

In the figure, the detector is indicated by a square box with a $D$ in it. Notice the detector is near the bottom leg located in the middle, so that if the atom goes through that leg, then the atom will emit a photon at $L/2$ which will trigger the detector, thus indicating the passage of the particle at $L/2$ down that leg.

The system states at each leg are shown in the figure. We use the following notation for the detector and atom: $\left|d_0, e1, e2, \gamma 1, \gamma 2\right\rangle$, where $d_0$ ($d_1$) stands for detector unactivated (activated); $e1$ ($g1$) stands for first mode of atom excited (in ground state) and $\gamma 1=1$ ($\gamma 2=0$) stands for one (no) photon from first (second) mode in vacuum; $\gamma 2=1$ stands for one photon from second mode in vacuum. Because the $\gamma$'s and the $e$'s are paired, we adopt a simplified notation of just showing the state of the EM field from mode 1 and 2, i.e. $\left|d_0, \gamma 1, \gamma 2\right\rangle$.

We multiply by the spin to account for it and by a wavefunction to account for the spatial part, so that the entry state is: $(|\uparrow\rangle + |\downarrow\rangle)\psi_r(x)\left|d_0 00\right\rangle$. The subscript $r$ ($u$) refers to rightward (upward) traveling wave; $rd$ ($ru$) refers to rightward traveling down (up) leg.

Now, note in the figure a fiber optic is inserted. With its given location, if the atom goes up (into the first leg) at the first mirror and therefore decays for the second time in the third leg (and first time in first leg, but this photon is not detected), the light it emits goes through the fiber optic and triggers the detector. Without this fiber optic, the state entering from the third leg would be $|\uparrow\rangle\psi_d(x)\left|d_0 11\right\rangle$, while the state in the lower leg (the leg defined by the beam that goes directly through the first mirror) is $|\downarrow\rangle\psi_{rd}(x)\left|d_1 11\right\rangle$, so that at the output one, has: $\left(|\uparrow\rangle\left|d_0 11\right\rangle + |\downarrow\rangle\left|d_1 11\right\rangle\right)\psi_{r2}(x)$. Hence, one does not recover an atom with rightward spin! Why? Because the two states up and down can no longer superpose. However, with the fiber optic in, one gets: $|\uparrow\rangle\psi_d(x)\left|d_1 11\right\rangle$ in the third leg and retains $|\downarrow\rangle\psi_{rd}(x)\left|d_1 11\right\rangle$ in the lower; so, we get: $(|\uparrow\rangle + |\downarrow\rangle)\left|d_1 11\right\rangle\psi_{r2}(x)$ and the spins again superpose![63] This is so even though we know which way the atom went, for when the atom goes upward, the lower detector triggers at $t = 5L/2/\mathrm{v}$, while in the bottom leg, it triggers at $L/2/\mathrm{v}$ (and the second photon is lost, not detected), where v is speed of atoms.[64]

---

[63] We could analyze further and speak of noting the detector entanglement with the brain, upon human observation. However, that only adds to the $d1$ ket a $b1$ label, i.e. an indication of the "brain detected state."

[64] To see that the requisite "which way" timing does not substantially affect the above state-ket description, consider the more full account of our detector and of our "gun" that shoots the atoms at the input mirror. Assume that when the detector is triggered by light from the atom traveling through the interferometer, the detector itself, then, emits a pulse of collimated light that exits the back of the detector. Namely, the pulse of light travels away from the apparatus, for example, in a (not-shown) fiber optic. Allow this light to travel far way from the apparatus. Indeed, wait until after the output of the interferometer has been clearly

Now for the collapse approach. In collapse, measuring the atom in the lower leg should get rid of the second leg wavefunction, completely collapsing it to a position state at the detector, once the detector triggers at $t = L/2/\mathrm{v}$. This means no interference. This contradicts the fact that there will be interference in the case with the fiber installed!

Note also another problem with collapse. One could draw it out using this example, but it suffices to say that Pearle has shown that collapse violates conservation of momentum/energy.[65]

## VII: Conclusion

This section VI example shows that the part of the wavefunction obliterated during the collapse process is actually still there and, thus, the standard (Copenhagen) interpretation leads to a prediction that radically disagrees with experiment. It shows we cannot localize the wavefunction simply by making a position measurement (forcing an *R* process).

In this paper more generally, we have seen, through a detailed analysis of the two slit experiment, that Hegerfeldt localization drives us reconsider and then (after a full analysis) reject the collapse as a postulate of ordinary QM.

And, in the process of this analysis, we were led to understand measurement in its most natural terms with the SE as the sole form of evolution of the wave-function whose square magnitude, in turn, is a probabilistic description of an ensemble of "quantum mechanically identically" prepared systems. Furthermore, each system in the ensemble is seen as composed of multiple entities (particle, guiding wave structure and stochastic agents), the wavefunction hinting at something about all three kinds of entities, including something of the nature of the guiding wave structure's[66] interference capacities. Indeed, in section III, four essential, but little known, points on the meaning and understanding of ordinary QM are given, which help us avoid the paradoxes and even contradictions that are generated when one tries to understand QM. The four points also help us avoid the erroneous predictions that occur in applying the quick pragmatic (collapse) understanding in certain circumstances. In particular, we see that collapse has a place as an approximation in which one considers a subset of a full ensemble and/or parts of each system in the full ensemble which both explains why the approximation works and where it doesn't. We do this by recalling that the wavefunctions that solves the SE carries information about the statistics of the full ensemble and in making our collapse approximation, we hew to parts we are interested in and ignore the rest. In this way, the

---

registered and then record the time of the pulse. This time is then compared with the time that the gun was fired, which is recorded by the gun each time it is fired. Since the subscript "1" on "*d*" used in our notation (for example, $|\downarrow\rangle \psi_{rd}(x)\left|d_1 11\right\rangle$) simply means that the detector is fired (and when "0", not fired), in our more specific case, it means that the pulse has started traveling down the fiber optic away from the apparatus.

[65] P. Pearle, *Wavefunction Collapse and Conservation Laws*. Foundations of Physics **30**, 1145–1160 (2000). https://doi.org/10.1023/A:1003677103804

[66] Note the guiding wave structure mentioned here is radically distinct from de Broglie Bohm theory; this is especially evident in the fact that no extra implicit QM principles are here added to allow prediction of single trajectories. This minimality is the key feature of the natural interpretation; QM is taken as it is, carefully avoiding any gratuitous addition of principles (through rules or otherwise) to ordinary QM to get, for example, a more predictive QM than we actually have. In it we let experiment and the SE-alone evolution of the wavefunction (which, for example, gives probability through the born rule or more generically by expectation values of operators) guide our understanding of its physical meaning.

natural (ensemble) interpretation arises through the analysis and explains the actual results simply.

Collapse to a position (or finite volume around a position) is obviously the first meaning of localization, and it is not allowed by QM. Here we saw this simply means that the wavefunction cannot be localized, and we also saw that this squares with requirements for proper solutions to the SE. This, in turn, means that the ensemble description given by the wavefunction is limited to those that cannot be localized. It doesn't mean we cannot say for other reasons (special relativity, for example) that the particle in a subensemble of systems is localized! It means you cannot make a prediction using ordinary QM, i.e. SE evolution of $\psi$ (which is all we know non-relativistically) that tells you that it will remain localized. Neither can you do it using the formalism of QFT, but QFT guarantees the particle cannot travel superluminally, so other predictions can be made. However, they will not involve giving sharp meaning to an evolving causal (SR) boundary to the confining region in which the particle is; we need the wavefunction-like thing for this and there is none in the full relativistic regime. Moreover, any operator that purports to give particle number in a given state, i.e. $\langle\psi|\hat{N}|\psi\rangle$, will have infinite tails. Again, we can make predictions but not evolve the localized number probability density.

This, in turn, makes it clear that the argument made against the existence of particles in QFT (requiring everything to be simply fields) is not valid. Viz, the statement that "one cannot localize the place of the particle to a finite place because it will immediately be anywhere, and, therefore, particles (which are, by definition, in a finite place) cannot exist in QFT" is incorrect. Through the Reeh-Schlieder theorem, we have seen that, although undetectable interactions can travel superluminally, the localized detections, which lead one to speak of particles, must travel at or less than the speed of light. Just because QFT does not have a sharp equivalent of a probability of finding a particle that can make predictions about how an ensemble of them behaves doesn't mean the localized detections aren't there and don't indicate localized objects (particles). In short, the natural inclusion of particles along with fields in QFT is upheld by this analysis.

We saw the distinction between pragmatic approximation (which is almost always necessary) and our understanding of that approximation. And, we saw, at root, the meaning of QM that is reflected in the equations, i.e. the interpretation that has been lost in the practice of a QM that we can make work "well enough." The years of handwringing on the Copenhagen interpretations' paradoxes and even absurdities are the tip the iceberg of the obviousness of that fallacy of confusing pragmatism with understanding.

Other more general examples of this over-quick pragmatic-induced certainty are at hand. Only recently did people realize that one photon can be absorbed by two separated atoms, effectively splitting the oft spoken of "quantum" of light."[67] The original Aharonhov-Bohm analysis[68] recently was shown to have not taken into account the full QFT of its three elements, which can make one come up with wrong answers for the result. But, we don't need examples to make us sure that if we don't know what we are doing, it will inevitably lead to wrong statements and even wrong predictions if carried on long enough.

[67] L. Garziano et al. "*One Photon Can Simultaneously Excite Two or More Atoms*", Phys. Rev. Lett. **117** No. 4, 043601 (2016). https://doi.org/10.1103/PhysRevLett.117.043601. Note this paper adds back the "counter-rotating terms" that were discarded in rotating wave approximation used in Jaynes-Cummings model.

In our current case, we have seen collapse leads to predictions that are completely wrong in addition to being fraught with problems which we have shown and, finally fundamental problems of existence of things recognized from early on by the QM founders like Einstein and Schrodinger.

Let's end by discussing the treatment of collapse in the very popular text book by Griffiths; it says: "What if I made a second measurement, immediately after the first? Would I get C again, or does the act of measurement cough up some completely new number each time? On this question everyone is in agreement: A repeated measurement (on the same particle) must return the same value. Indeed, it would be tough to prove that the particle was really found at C in the first instance, if this could not be confirmed by immediate repetition of the measurement."[69,70] In the following footnote, one can note a similar, "here's what seems to work" approach in Dirac.[71,72]

This is obviously false, for all that is necessary is that the measurement readout (digital or needle) be correlated with the position (e.g.) after the measurement is finished. It is not necessary that it finish in an eigenstate, for the measurement itself can in principle change the original value.

As described formally and succinctly in the introduction, we have seen generically that one-to-one correlation between the system to be measured and the measuring apparatus pointer is sufficient for a measurement.[73]

---

[68] P. Pearle, A. Rizzi, *Quantized Vector Potential and Alternative Views of the Magnetic Aharonov-Bohm Phase Shift*, Phys. Rev. A 95, 052124 (2017). P. Pearle, A. Rizzi, *Quantum Mechanical Inclusion of the Source in the Aharonov-Bohm Effects*, Phys. Rev. A 95, 052123 (2017).

[69] D. J. Griffiths, and D. F. Schroeter, *Introduction to Quantum Mechanics*, 3rd edition, (Cambridge University Press, 2018.).

[70] A few sentences later it defines locational collapse saying: "We say that the wave function **collapses**, upon measurement, to a spike at the point C (it soon spreads out again, in accordance with the Schrödinger equation, so the second measurement must be made quickly)."

[71] Dirac, P. A. M. *The Principles of Quantum Mechanics*, 4th edition, Oxford: University Press, 1958, pg 36.

[72] "When we measure a real dynamical variable $\xi$, the disturbance involved in the act of measurement causes a jump in the state of the dynamical system. From physical continuity, if we make a second measurement of the same dynamical variable $\xi$ immediately after the first, the result of the second measurement must be the same as that of the first. Thus after the first measurement has been made, there is no indeterminacy in the result of the second. Hence, after the first measurement has been made, the system is in an eigenstate of the dynamical variable $\xi$, the eigenvalue it belongs to being equal to the result of the first measurement. This conclusion must still hold if the second measurement is not actually made. In this way we see that a measurement always causes the system to jump into an eigenstate of the dynamical variable that is being measured, the eigenvalue this eigenstate belongs to being equal to the result of the measurement."

[73] A reviewer of a previous measurement paper asserted, without reference, that there are three requirements for measurement: a) interaction between system and measuring apparatus b) irreversible record result c) making use of the information to update the representation of the system. Clearly, "a" is necessary and "c" is normally done but not necessary. One has a measurement before one makes use of it. "b" is certainly not necessary and it is problematic. There is no proof that there are, *in principle*,

irreversible interactions in QM as it currently exists (indeed, the SE is time reversible), let alone that they must exist in every measurement. A fortiori, adding "b" begs the question. Namely, adding a condition of irreversibility in all measurement activities is tantamount to adding something like an *R*-process to QM.